\documentclass[superscriptaddress,article,pra,showpacs,nofootinbib,10pt, twocolumn]{revtex4-1}
\usepackage{amsmath,xcolor,amsfonts,amsthm,amssymb,graphicx, srcltx,hyperref}
\usepackage[normalem]{ulem} 
\usepackage{tikz}
\usepackage{acronym}

\newcommand{\thr}{{V_{\rm T}}}          	
\newcommand{\mean}[1]{m_{#1}}          	
\newcommand{\var}[1]{v_{#1}}        	
\newcommand{\varLogn}[1]{\sigma^2_{#1}}    	
\newcommand{\nSiftTh}[1]{S(#1)}       	
\newcommand{\eSiftTh}[1]{E(#1)}           	
\newcommand{\iPktTh}[1]{\mathcal{I}(\thr)}      	
\newcommand{\nPktTh}[1]{N_{\rm{P}}(#1)}      	
\newcommand{\nSiftPktTh}[1]{\overline{\mathcal S}(#1)} 	
\newcommand{\erf}[1]{{\rm{erf}}\left(#1\right)}      	
\newcommand{\qber}{Q}              	
\newcommand{\qberTh}[1]{\qber(#1)}       	
\newcommand{\optThrTh}{V_{T,\rm{opt}}^{(\rm{th})}}   
\newcommand{\Nb}{n_b}	
\newcommand{\blu}[1]{{\color{blue}#1}}

\renewcommand{\blu}[1]{}

\begin{document}

\title{Adaptive Real Time Selection for Quantum Key Distribution in 
Lossy and Turbulent Free-Space Channels}

 \author{Giuseppe Vallone}
 \author{Davide Marangon}
 \author{Matteo Canale}
 \author{Ilaria Savorgnan}
 \author{Davide Bacco}
 \affiliation{Department of Information Engineering, University of Padova, via Gradenigo 6/B, 35131 Padova, Italy}
 \author{Mauro Barbieri}
 \affiliation{Department of Physics and Astronomy, University of Padova, vicolo dell'Osservatorio 3, 35122 Padova, Italy}
 \author{Simon Calimani}
 \affiliation{Department of Information Engineering, University of Padova, via Gradenigo 6/B, 35131 Padova, Italy}
 \author{Cesare Barbieri}
 \affiliation{Department of Physics and Astronomy, University of Padova, vicolo dell'Osservatorio 3, 35122 Padova, Italy}
 \author{Nicola Laurenti}
 \author{Paolo Villoresi$^*$}
 \affiliation{Department of Information Engineering, University of Padova, via Gradenigo 6/B, 35131 Padova, Italy}

\email [Correspondence and requests for materials should be addressed to P.V.]{paolo.villoresi@dei.unipd.it}

\date{\today \   \  \  \ $^*$e-mail: paolo.villoresi@dei.unipd.it}

\begin{abstract}
The unconditional security in the creation of cryptographic keys obtained by Quantum Key Distribution (QKD) 
protocols will induce a quantum leap in free-space communication privacy in the same way that we are beginning 
to realize secure optical fibre connections. 
However, free-space channels, in particular those with long links and the presence of atmospheric turbulence, are
affected by losses, {fluctuating transmissivity} and background light that impair the conditions for secure QKD. 
Here we introduce a method to contrast the atmospheric turbulence in QKD experiments. 
Our Adaptive Real Time Selection (ARTS) technique at the receiver
{is based on the selection of the intervals with higher channel transmissivity.}
We demonstrate, using data from the Canary Island 143 km free-space link, that conditions with unacceptable average QBER
which would prevent the generation of a secure key can be used once parsed according to the instantaneous scintillation using the ARTS technique.
\end{abstract}

\maketitle

\section{Introduction}
The transmissions of quantum states to a distant receiver, which may be placed on a mobile terminal or on-board of an orbiting station, is the frontier of Quantum Communications (QC) and of QC protocols that are based on the transmission and detection of quantum bits, or qubits. 
As such, it has been investigated by different groups ~\cite{vane98sci,benn99sci,lo99sci,hugh11sci} as well as included in continental roadmaps for technology development, as in the case of Europe \cite{roadmapEU}, Japan \cite{roadmapJP} or China \cite{xin11sci}.

In free-space communications, background photons and detector noise are unavoidable sources of quantum bit error rate (QBER), 
which limits the range of quantum protocols. 
Indeed, in the case of Quantum Key Distribution (QKD), an unconditional secret key may be generated only if the QBER is 
below a given threshold. For this reason, 
free-space QKD demonstrations have so far been realized typically during dark nights 
or by using very narrow spectral filters that impose a low key rate already on urban scale
\cite{jaco96opl,butt98prl,hugh02njp,nord02spie,else09njp,peev09njp,fedr09nap,garc13apo,peun14prl, vall14prl}. 
To overcome this limitation,
a modeling of the probability distribution for the transmission coefficient was exploited, in order to devise a post-selection technique for slowly fluctuating channels \cite {seme09pra,seme10pra}.  
The restriction to the analysis of the sifted bit rate on a millisecond time scale was inspired by this approach \cite{erve12njp}.  
However, in the case of strong turbulence, due to the inherent high losses 
the channel transmissivity cannot be reliably estimated in the intrinsic time scale of its variation by the received photons. 

A different approach is found in the CAD1 and CAD2 distillation schemes  \cite{bae07pra}, that represent a generalization of Maurer's advantage distillation technique \cite{maur93ieee}. Sequences of correct (possibly non consecutive) sifted bits are joined together and one single secure bit is distilled out of each sequence. The length of each sequence should be chosen according to a trade-off scheme, because longer sequences allow to distil keys with higher channel QBERs, but provide a lower key rate in the case of low QBERs. However, in a turbulent, rapidly time-varying channel,  the effectiveness of such solutions would be limited by the difficulty of choosing the suitable parameters of the distillation strategy according to the varying QBER. Another generalization of the advantage distillation in \cite{maur93ieee} was proposed in \cite{wata07pra}: parities for many pairs of bits are shared between Alice and Bob along the public channel. Those pairs with non-matching parities are discarded, while the remaining ones (over which the QBER is lower) are syndrome-decoded.
However, the above presented distillation methods do not take advantage of the intrinsic QBER variability of the channels, as they rely on the assumption that the channel maintains its QBER stable long enough to allow optimization of their parameters.


\begin{figure*}[t!]
\centering
\includegraphics[width=0.9\textwidth]{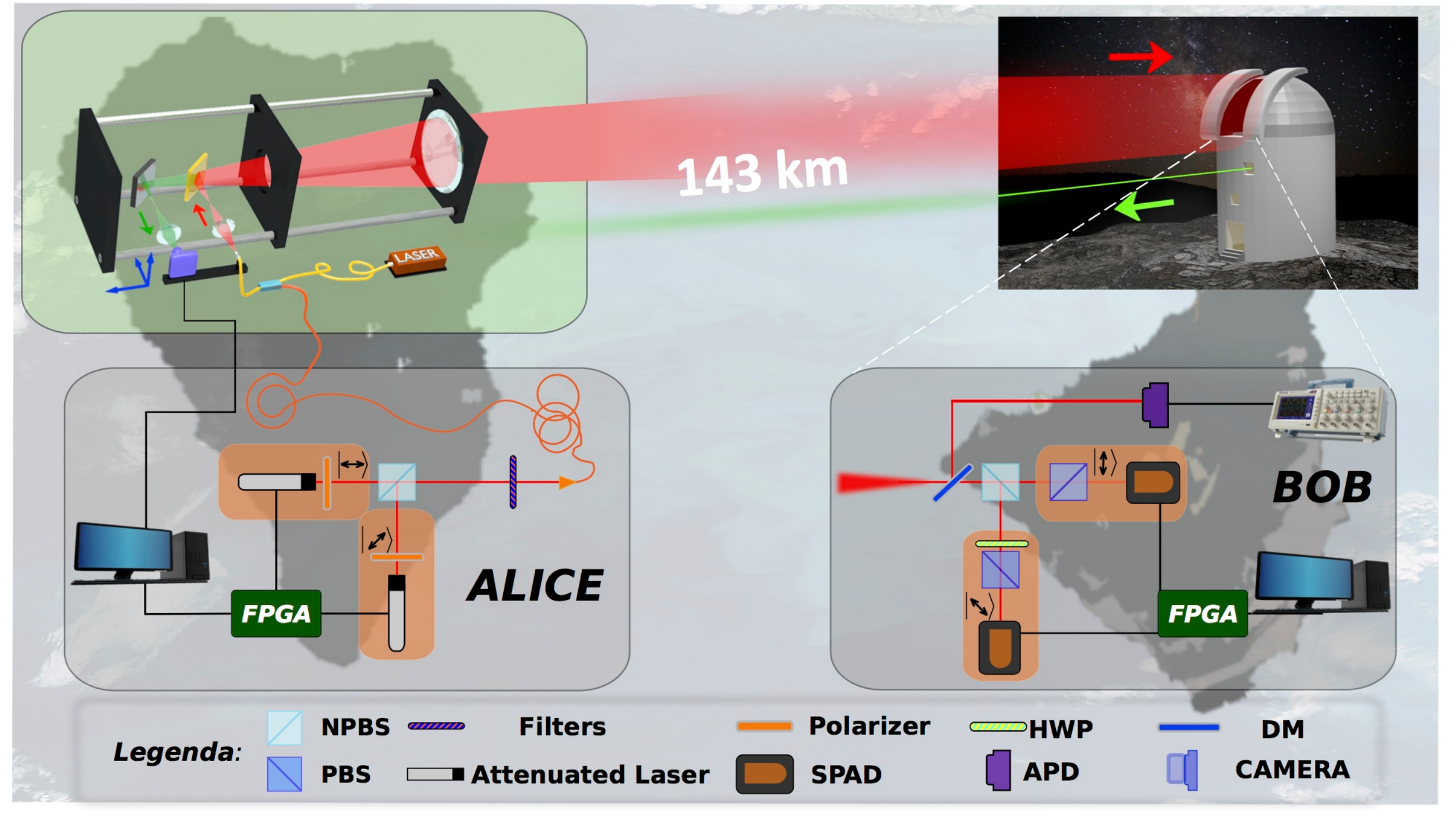}
\caption{(color online) Experimental setup: Alice is located at JKT observatory in La Palma. 
Qubits are prepared using a FPGA-controller, linear optical components and attenuated lasers.  
Alice telescope is also configured to acquire the beacon beam sent by Bob, located at the Optical Ground Station in Tenerife, 
and used for tracking the transmitter in order to stabilize the pointing. }
\label{fig:setup}
\end{figure*}
Here, we devise a method that exploits strong atmospheric turbulence for secret key generation, in conditions in which the 
long-time average QBER is too high for secure communication. 
The temporal profile of the transmissivity in a long and strongly turbulent channel 
has characteristic peaks lasting few milliseconds, and following a lognormal distribution  \cite{capr12prl}. 
On these grounds, we will introduce and demonstrate an adaptive real time selection (ARTS) scheme, inspired by the experimental observations reported  in  \cite{capr12prl}, which were subsequently analyzed numerically using the split-step method in   \cite{feng13col}.
The scheme is based on the estimation of the link transmissivity over its intrinsic time scale 
by an auxiliary classical laser beam, copropagating with the qubits, but conveniently interleaved in time. 
In this way, the link scintillation is monitored in real time and the high channel transmissivity intervals 
corresponding to a viable QBER for a positive key generation rate can be selected.
We will present a demonstration of this protocol in loss conditions that are equivalent to long distance and satellite links, 
and with scintillation range corresponding to moderate to severe weather. 

\section{Experimental setup}         
The  143 km free-space channel between La Palma and Tenerife Islands 
was used as the best available testbed for turbulence and long links, as sketched in Fig.  \ref{fig:setup}. 
The transmitter (Alice) was located at the JKT observatory in the island of La Palma. 
At the transmitter side, qubits were made by using strongly attenuated $850$ nm lasers. 
In the same location we also used a 30 mW classical laser beam (probe) at 808 nm to estimate the link transmissivity
using pulses of $100\ \mu$s duration at $1$ kHz repetition rate. 
The encoding of the quantum signal was then obtained by controlling the lasers with an FPGA.
Classical and quantum beams were coupled into single mode fibers and injected into a fiber beam coupler. The output was sent to a Galilean telescope that we designed specifically with  a singlet aspheric lens of 230 mm diameter and 2200 mm of focal length as the main component. 
The beam diameter size of about 20cm provided the required class-1M eye-safe conditions for the free-space propagating beams. 
In vacuum, the telescope would produce, after 143 km of propagation, a spot comparable to the primary mirror of the receiving telescope, in order to maximize the power transfer between the two parties. To compensate the beam wandering induced by the atmosphere, we implemented a feedback loop for controlling the transmitting direction: the fiber delivering the signal to the transmitter was mounted on a XYZ movable stage placed close to the focal place of the 230mm lens, with computer controlled stepped motors. On this same stage, we mounted a CCD sensor which acquired a green (532 nm) beacon laser sent by Tenerife toward Alice telescope. The camera is placed in order to measure an image of the singlet focal plane: the wandering of the beacon on the CCD was then analyzed in real time by a software that moves the XYZ stage to compensate the movement of the beacon spot on the camera.

At the receiver part (Bob), in Tenerife, we used the $1$ m aperture telescope of the ESA Optical Ground Station to receive the signals. After the Coud\'e path, we collimated the beam and we divided the classical and quantum signal by using a dichroic mirror. The qubits were measured in two bases and the counts detected by the two single-photon avalanche photodiodes(SPAD) were stored on a FPGA. The probe beam is detected by an high-bandwidth APD (avalanche photodetector) and then registered and stored by an oscilloscope.

In order to measure the QBER of the channel, we used the data structure optimized for  free-space QKD implementation based on a recent implementation of the B92 protocol~\cite{benn92prl,cana11isa}. A raw key is composed by N packets of 2880 bits each, 
sent at the rate of 2.5 MHz; as regards the payload slots, Alice sends two qubits separated by 200 ns. Due to communication 
constraints with the FPGA, each packets is sent every 20ms resulting in an average sending rate of 150 kHz.
The two FPGAs are synchronized every second by a pulse-per-second (pps) signal provided by two GPS receivers located in the two islands.

We point out that, at the transmitter side, the pulses contain on average more than one photon, while at the receiver side we work in the single photon regime. Our aim, in fact, was to simulate a possible realistic scenario where one would employ fast (hundreds of MHz to GHz) free-space QKD systems which are nowadays commonly available. Since our system has a transmission rate of 2.5 MHz, the detected rate is comparable to the rate observable with a transmitter emitting true single photon pulses with a repetition rate of about 1GHz, assuming that the amount of optical and atmospheric attenuation is  fixed.

\begin{figure}[t!!]
\centering
\includegraphics[width=8.5cm]{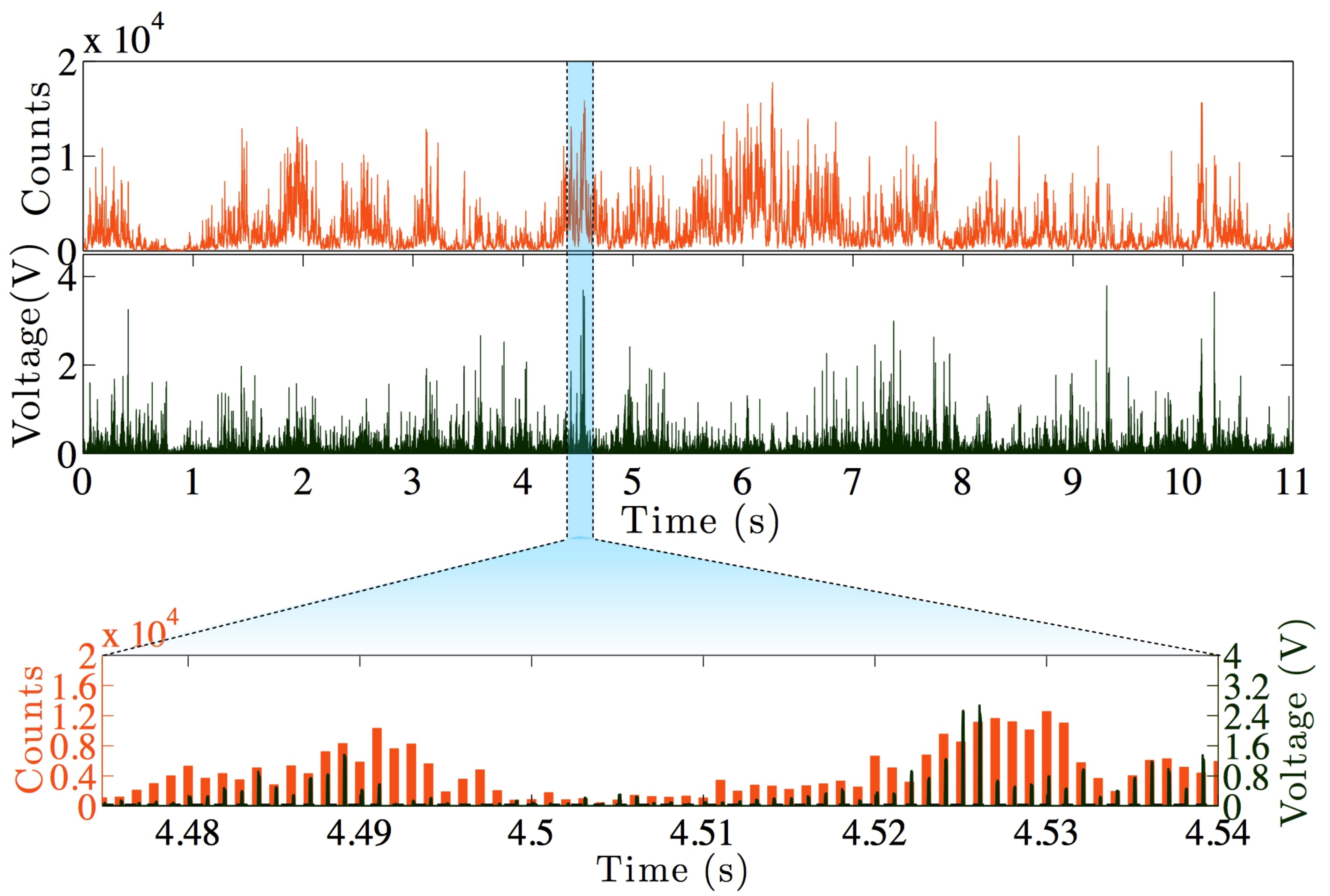}
\caption{(color online) Comparison between the counts detected by the SPAD (green dark line) and the voltage measured by the fast photodiode at the receiver (red light line). In the inset we show a zoomed detail of the acquisition (between $4.48s$ and $5.54s$) in order to better appreciate the correlation between the quantum and classical signal. We chose a particular inset but in all the acquisition the two signals are correlated.}
\label{correlation}
\end{figure}

\section{Realization of the ARTS method}
In order to assess the correspondence between the intensity of the probe beam and the
number of photons received on the quantum channel, the link transmissivity was estimated with a fast photodiode on the probe signal.
The single photon stream was detected by a Single Photon Avalanche Photodiode (SPAD) and binned in 1 ms intervals.
With this measurement we demonstrate  a  good correlation between the two signals, 
as shown in Figure \ref{correlation} for an acquisition time of 11 seconds.

To further demonstrate the correlation we performed the ARTS method, consisting in the following procedure: given a set of $L$ packets (each of $1ms$ length), we denote by $V_i$ the probe signal amplitude and by $S_i$ 
the number of detected photons in the quantum signal for the $i$-th packet. 
We set a threshold value $\thr$ for the probe voltage and post-select only those packets such that $V_i > \thr$; in particular, we denote by $\iPktTh{\thr} = \{i : V_i > \thr\}$ the indexes of the packets for which the above condition holds and by $\nPktTh{\thr}$ the corresponding number of packets, that is, $\nPktTh{\thr} = |\{\iPktTh{\thr}\}|$. Furthermore, we define the following quantities:
\begin{eqnarray}\nSiftTh{\thr} &=& \sum_{i \in \iPktTh{\thr}} S_i, \quad \nSiftPktTh{\thr} = \frac{\nSiftTh{\thr}}{\nPktTh{\thr}}\end{eqnarray}
with $\nSiftTh{\thr}$ representing the total number of detected bits and $\nSiftPktTh{\thr}$ the mean number 
of detection per packets after the post-selection performed with threshold $\thr$.

\begin{figure}[t!]
\centering
\includegraphics[width=0.45\textwidth]{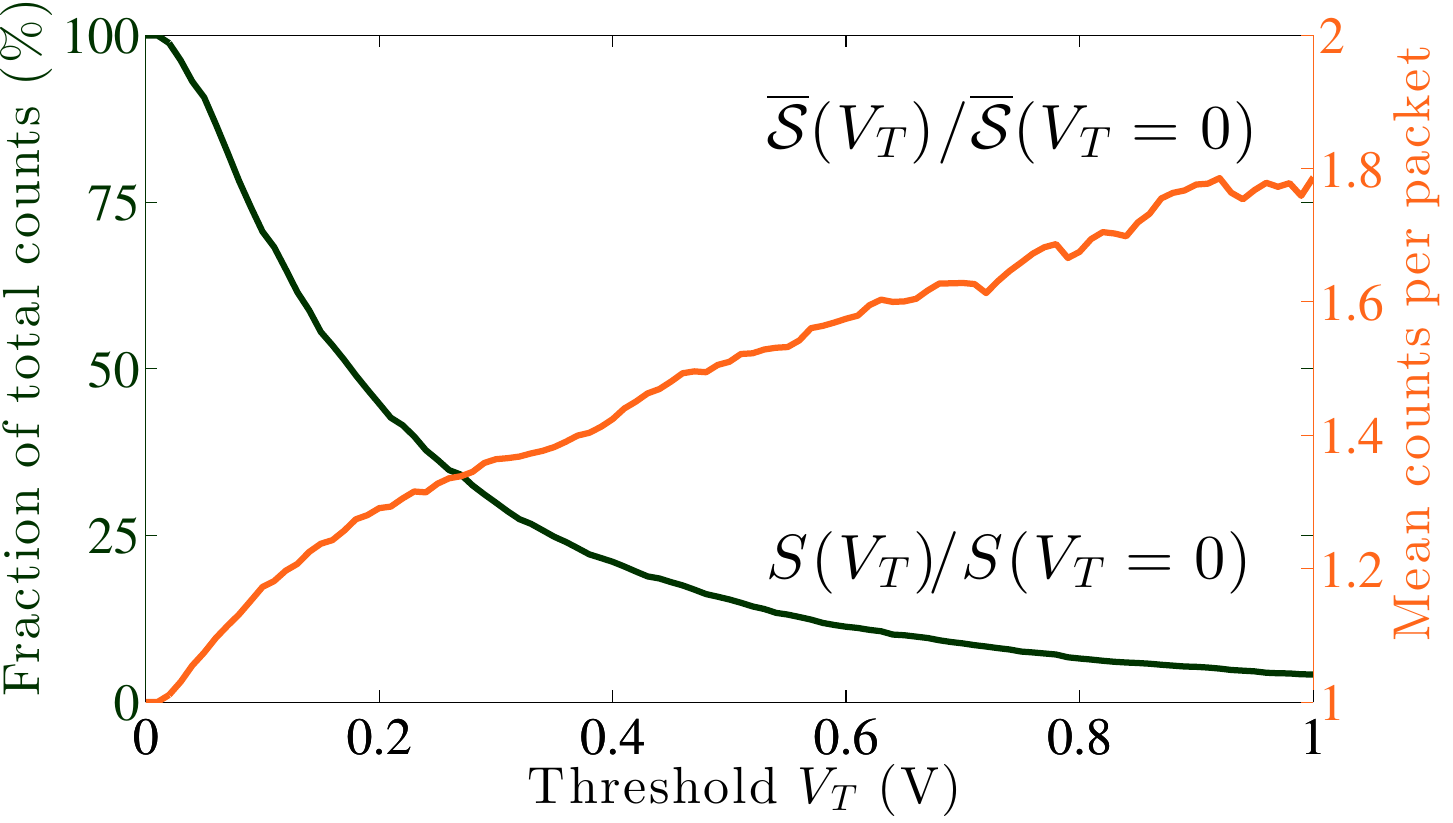}
\caption{(color online) Mean counts per packet $\nSiftPktTh{\thr}$ (normalized to the mean counts obtained without thresholding) and fraction of total count $\nSiftTh{\thr}/\nSiftTh{\thr=0}$ in function of the probe threshold.}
\label{fig:threshold_1}
\end{figure}

The effect of the ARTS procedure can be clearly appreciated in Fig. \ref{fig:threshold_1}, where $\nSiftPktTh{\thr}$ (normalized to the mean counts obtained without thresholding) is plotted (green line) as a function of the threshold: a higher threshold value corresponds 
to a larger mean number of counts per packet. 
This demonstrates that the probe and quantum signal are correlated and one can significantly improve the signal-to-noise ratio (SNR) 
by thresholding. Here we define the SNR as the ratio between the overall signal (true signal and background) and the background. As side effect, we observe that the pre-selection also decreases the overall number of detections in the transmission $\nSiftTh{\thr}$ as can be noticed by considering the ratio $\nSiftTh{\thr}/\nSiftTh{\thr=0}$ (blue line).

We then apply the results previously described to a QKD experiment. In particular, we show that, the thresholding gives, in some cases, benefits in terms of the secret key length, even if the total number of sifted bits decrease. Indeed, when the QBER is above the maximum value tolerable for QKD ($11$ \% for the BB84 protocol) is not possible to produce secure keys. However, by the ARTS method we will reduce the QBER below 
such limit, allowing secure key generation. We point out that at the receiver the beam has a mean photon number per pulse below $1$, namely it is the single photon level. 

First, given the number of errors $E_i$ in the $i$-th packet, we define the overall number of errors $\eSiftTh{\thr}$ and the quantum bit error rate $\qberTh{\thr}$ in the post-selected packets as
\begin{equation}
\eSiftTh{\thr}= \sum_{i \in \iPktTh{\thr}} E_i\,,
\quad
\qberTh{\thr} = \frac{\eSiftTh{\thr}}{\nSiftTh{\thr}}.
\end{equation}

For evaluating the impact of the ARTS procedure on the performance of a quantum key distribution system, it is important to study the two complementary effects of thresholding: on one side, an higher threshold  increase the mean detected bits per packet $\nSiftPktTh{\thr}$. 
On the other side it decrease the total detections $\nSiftTh{\thr}$. Both effects influence the achievable secret key rate of the system, and an optimal trade-off should be found.

\begin{figure}[t]
\centering
\includegraphics[width=0.49\textwidth]{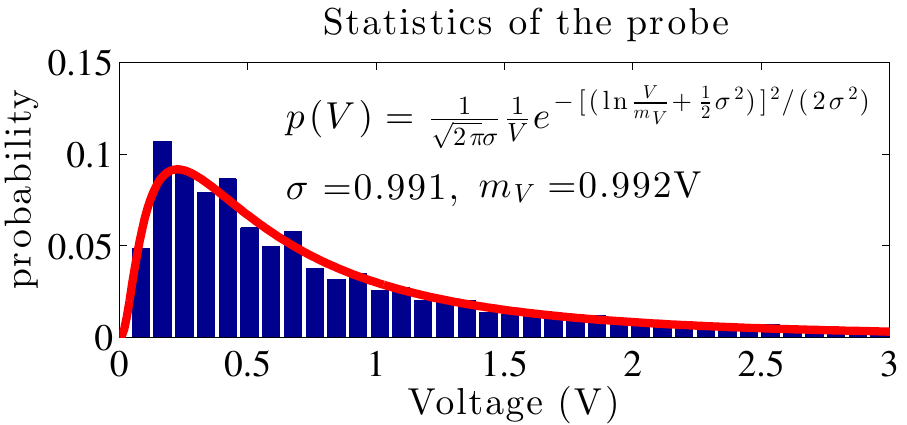}
\caption{(color online) Experimental occurrences of probe intensities (measured by photodiode voltages) and lognormal fit.}
\label{fig:lognormal}
\end{figure}

We first derive the expected number of sifted bits and their bit error rate after thresholding.
As demonstrated in \cite{capr12prl}, the statistics of the transmission of a long free-space channel follows a log-normal distribution. The measured probe voltage at the receiver, being the transmitted intensity constant, follows the same distribution, given by $p(V;\mean{V},\varLogn{}) = \frac{1}{\sqrt{2\pi} \sigma}\frac1V e^{-[(\ln\frac{V}{\mean{V}}+\frac{1}{2}\varLogn{})]^2/(2\varLogn{})}$. In the previous expression
$\varLogn{}$ is related to the mean $\mean{V}$ and the variance $\var{V}$ of the probe intensities distribution by
$\varLogn{} = \ln\left(1+\frac{\var{V}}{\mean{V}^2}\right)$. As an example, we show in figure \ref{fig:lognormal}
the distribution of the  probabilities of occurrence of different photodiode voltages corresponding to different probe intensities
as measured on the focal plane of the receiver. The shown data are the same used in Figure \ref{correlation}.
According to the theory \cite{milo04job,capr12prl}, the probabilities follow a log-normal distribution, as demonstrated
by the continuous (red) curve representing the lognormal fit.

\begin{figure}[t!]
\centering
\includegraphics[width=0.49\textwidth]{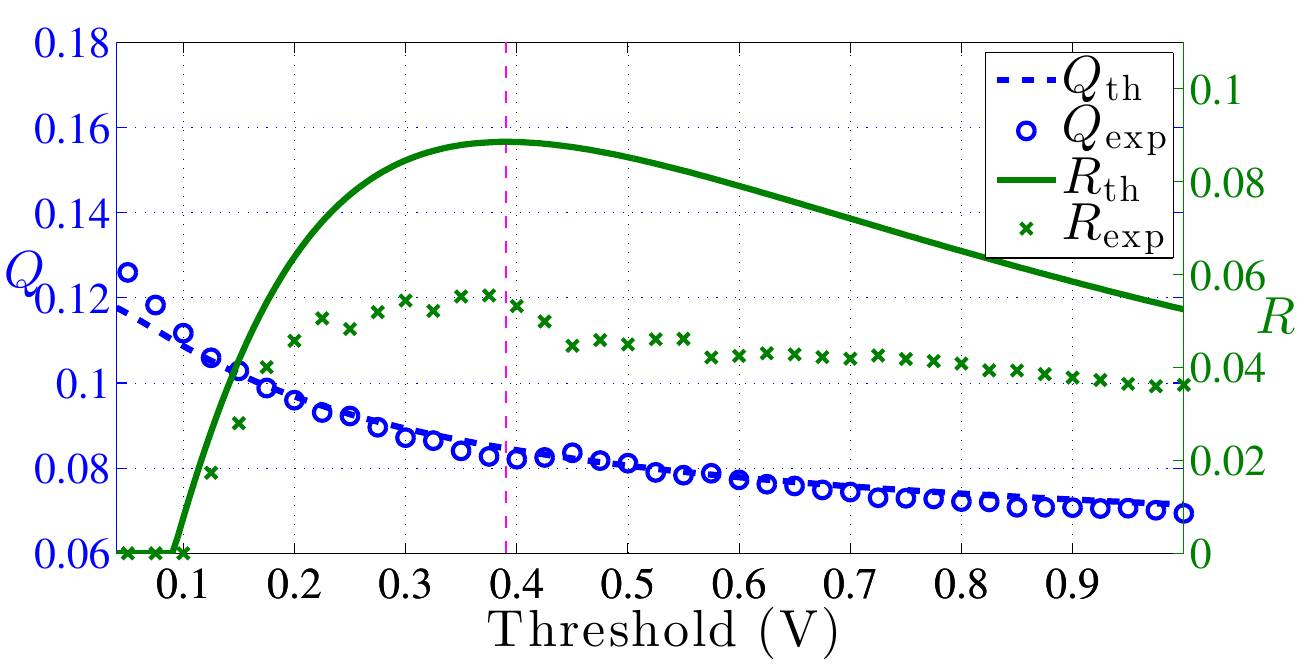}
\caption{(color online) Experimental QBER ($Q_{\rm exp}$) and secure key rate ($R_{\rm exp}$) in function of the probe threshold (measured by the photodiode voltage). With dashed and solid lines with respectively show the corresponding theoretical predictions ($Q_{\rm th}$ and $R_{\rm th}$).}
\label{fig:threshold_2}
\end{figure}

In the following analysis, we assume that the number of detected photons and the probe intensity 
have completely correlated log-normal distributions \cite{capr12prl}.
This hypothesis implies that both distributions have the same parameter $\varLogn{}$.
Then, we can predict the number of packets above threshold $\nPktTh{\thr}$ and
the number of sifted bits surviving the thresholding $\nSiftTh{\thr}$ in case of null background as 
$\nSiftTh{\thr} /\nSiftTh{0}=\int^{+\infty}_{\thr}\frac{V}{\mean{V}}p(V;\mean{V},\sigma)d V$ and
$\nPktTh{\thr}/\nPktTh{0}=\int^{+\infty}_{\thr}p(V;\mean{V},\sigma)d V$.
By taking into account the background clicks we get:

\begin{equation}
\label{eq:exp_nsift_th}
\begin{aligned}
\nPktTh{\thr}&= \nPktTh{0} \frac{1}{2}
\left[1-\erf{\frac{\ln \frac{\thr}{\mean{V}}+\frac{1}{2}\varLogn{}}{\sqrt{2\varLogn{}}}}\right]\,.
\\
\nSiftTh{\thr} &= \Nb N_P(\thr)+\frac{1}{2}[\nSiftTh{0}- \Nb N_P(0)]\times
\\
&\qquad\qquad\left[1-\erf{\frac{\ln \frac{\thr}{\mean{V}}-\frac{1}{2}\varLogn{}}{\sqrt{2\varLogn{}}}}\right],
\end{aligned}
\end{equation}
%
where $\Nb$ is the average background count per packet.
Indeed, experimental data suggest that the hypothesis of complete correlation between quantum and probe signal is not strictly satisfied, and the
previous expression turns out to be an approximation of the measured values. Still, 
they allow to derive a post-selection threshold that  improves the secure key rate, as shown in the following (e.g., in figure \ref{fig:threshold_2}).

We now define a further predictive model for estimating the bit error rate on the quantum channel as a function of the probe threshold. 
Let us assume that the average bit error rate on the quantum channel is $\mean{\qber}$ and that 
the number of counts per packet due to background noise is $\Nb $. Now, since background photons are not polarized, 
the corresponding bit error rate is $1/2$, and we can write the predicted quantum bit error rate $Q_{th}$ as a function of the threshold $\thr$, 
namely,
\begin{equation}\label{eq:exp_qber_th}
Q_{th}(\thr) = \mean{\qber} \left(1-\frac{\Nb }{\nSiftPktTh{\thr}}\right) + \frac{1}{2}\frac{\Nb }{\nSiftPktTh{\thr}}
\end{equation}
where $\nSiftPktTh{\thr}=\frac{\nSiftTh{\thr}}{\nPktTh{\thr}}$ is the predicted value for 
for the number of bits surviving thresholding.
Given these quantities, the asymptotic key rate of a QKD system based on the BB84 protocol\cite{benn84ieee} and the ARTS procedure (namely the probe thresholding mechanism) reads as follows:
\begin{equation}\label{eq:exp_rate}R(\thr) = \frac{\nSiftTh{\thr}}{\nSiftTh{0}}\left[1 - 2 h_2\left(\qberTh{\thr}\right)\right]\end{equation}
It is worth noting that using the asymptotic rate instead of the finite-lenght one \cite{bacc13nco, toma12nco}, 
may be considered a restrictive approach, especially because the post-selection further reduces the number of available sifted bits. 
However, it is sufficient to choose the size of the blocks before key distillation (i.e., information reconciliation and privacy amplification) 
large enough such that, without loss of generality, the asymptotic bound provides a reasonable approximation of the actual rate.

In figure \ref{fig:threshold_2}, we finally compare the theoretical (solid lines) and the experimental values (circles and crosses) 
of the measured QBER and the asymptotic key rate as a function of the probe intensity threshold in a data acquisition. 
The theoretical curves for the QBER and for the key rate were obtained by substituting in eq. \eqref{eq:exp_qber_th} and in eq. \eqref{eq:exp_rate} 
the estimates for the log-normal parameters $\mean{V}$ and $\varLogn{}$ of the probe signal distribution. 
The other two parameters, $\nSiftTh{0}$ and $\nPktTh{0}$, needed for predicting $\nSiftTh{T}$ and $\nPktTh{T}$, 
are directly measured (they correspond to the total sifted bits and the total number of packets received respectively).

The data shown in figure \ref{fig:threshold_2} correspond to an acquisition of $5 \cdot 10^5$ sifted bits in condition of high background, 
simulated by a thermal light source turned on in the receiver laboratory. The intensity of the background was chosen in order to obtain 
a mean QBER larger than $11\%$. In particular, we measured an average value of $\Nb  = 35.17$ for the background clicks per packet 
and we assume $\mean{Q} = 5.6 \cdot 10^{-2}$. As clearly shown in the figure, eq. \eqref{eq:exp_qber_th} provides a good approximation of the experimental curve.

Figure \ref{fig:threshold_2} also shows that there is a strong correspondence between the shape of the theoretical rate, $R_{th}$, and the measured rate, $R_{exp}$. The fact that the experimental points do not fit the expected curve can be ascribed to the
discrepancy in the empirical joint distribution of probe intensities and counts with respect to the model; 
in particular, we measured the following fitting parameters for the normalized log-normal distributions: $\varLogn{V} = 0.967$ for the 
probe intensities and $\varLogn{S} = 0.716$ for the photon signal. However, the derivation of the optimal threshold for maximizing 
the secret key length (magenta dashed line) from the probe distribution yields the optimal $\thr$ also for the experimental data. 
In particular, the optimal threshold inferred from the probe distribution is $\optThrTh = 375$ mV, 
and coincides with the one resulting from optimization on the experimental data, yielding a rate of $R(\optThrTh) = 5.55 \cdot 10^{-2}$.

We observe that, in the case of $\thr < 70$ mV, it is not possible to generate a secure key, being the QBER higher than the theoretical maximum 
(i.e., $\qber = 11 \%$). By increasing the threshold value above $70mV$ a non-zero secret key rate is achieved. With the optimal threshold value, 
the measured QBER is $\qberTh{\optThrTh} = 8.38 \cdot 10^{-2}$; 
a significant improvement with respect to the initial value, $\qberTh{0} = 13.14 \cdot 10^{-2}$, is therefore achieved. 
Finally, we observe that for increasing values of $\thr > \optThrTh$ the QBER still decreases, but so does the rate, 
since the reduction in the residual number of sifted bits does not compensate the advantage obtained from the lower QBER.
This result is of absolute practical relevance, as it shows that leveraging the probe intensity information is an enabling factor for quantum key distribution, allowing to distill a secret key.

As for the security of this post-selection approach as applied to a QKD system, 
no advantage is given to a potential attacker in the true single photon regime, being the thresholding nothing but a further sifting 
step on the received bits \cite{bae07pra,wata07pra}. 
If the attacker tried to force Alice and Bob to post-select a particular bit, in fact, she would alter the probe signal \emph{before} the disclosure of the preparation bases on the public channel, and, therefore, before she could actually know if her measured bit is correct. 
On the other hand, altering the probe statistics or interrupting the probe transmission would not yield any advantage to the attacker, as it would just break the correlation between the quantum and the classical signal and would thus result in a denial of service attack.
The security analysis gets more involved if we allow \emph{photon number splitting} (PNS) attacks. In that case, the attacker may force Bob to receive just the qubits for which the PNS attack was successful, i.e., only those pulses with multiple photons. A decoy state protocol may counteract this strategy, but its effectiveness with a turbulent and loss varying free-space channel has to be investigated.

\begin{figure*}[t!!!!]
\centering
\includegraphics[width=0.9\textwidth]{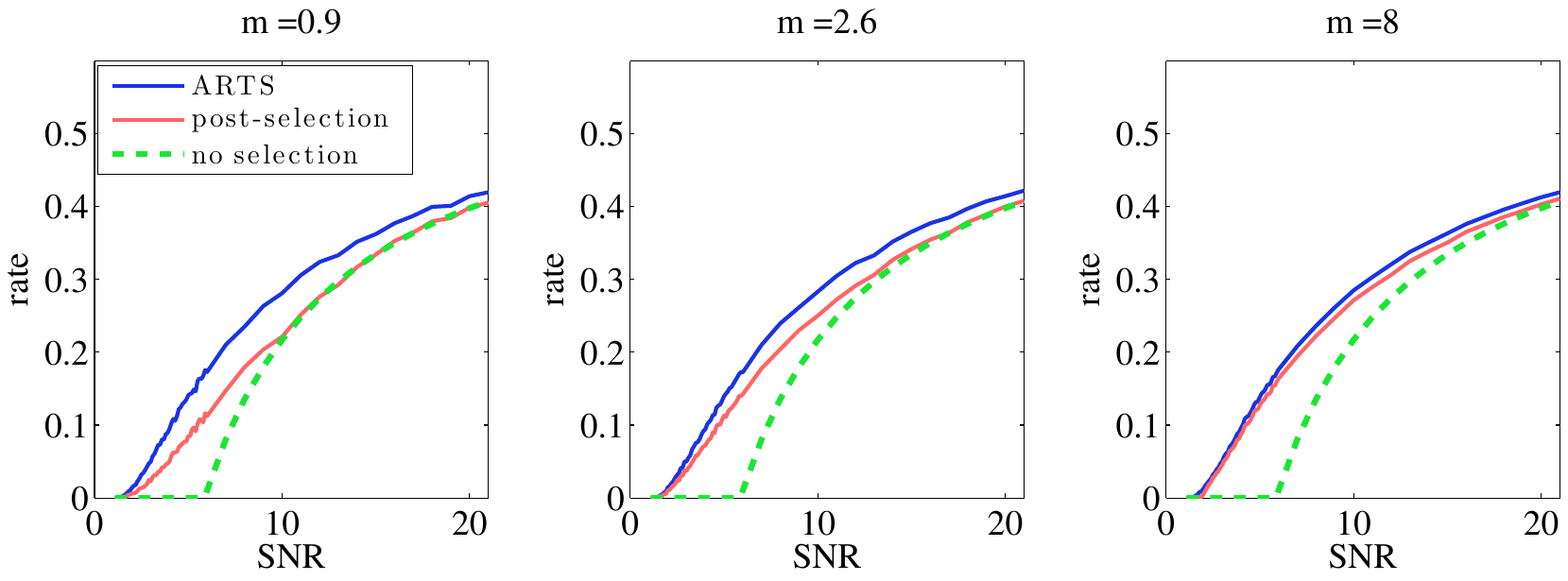}
\caption{(color online) Comparison between the rates achievable by the ARTS, the post-selection and the standard QKD technique (no selection). We assumed that the channel QBER is $3\%$ and the lognormal parameter is $\sigma=1$, similar to the parameter we measured in the tested free-space channel. 
The parameter $m$ represents the mean received bits per coherence time of the channel.}
\label{fig:comparison}
\end{figure*}
\section{Comparison with the post-selection on received bits}
The ARTS method can be compared with the technique introduced in \cite{erve12njp}, where a post-selection is performed when 
the number of received bits is above a given threshold.
The post-selection is effective only when the threshold is set in order to get at least several bits for coherence time of the channel: 
in fact, only in this condition it is possible to post-select the correct instants of high transmissivity. 
{By coherence time of the channel we denote the time (typically of the order of few milliseconds)
in which the transmissivity of the channel  can be considered constant}.
In the case of very turbulent channel and extreme environmental conditions (say mist or high humidity), 
the number of received bits per channel coherence time may not exceed the value of 10: 
in this case, the post-selection cannot be implemented and only the ARTS method remains effective.
This results is confirmed by the following analysis.
We performed a simulation to compare the two techniques by assuming that the probe and the signal statistics are perfectly correlated. 
The rate achievable in the two cases are shown in figure \ref{fig:comparison}, demonstrating that the ARTS methods always outperform 
the post-selection on the received sifted bits and it is particularly effective 
when the number of mean sifted bits received per coherence time of the channel are below $\sim10$ and the SNR is below 20.

\section{Conclusions}
In conclusions, we demonstrated a proof of principle of a method 
{that mitigates the detrimental effects of the atmospheric turbulence in QKD. The method exploits} 
the fluctuating transmissivity of the channel for allowing QC where not possible with standard approaches. 
The ARTS method is easily integrable in QKD systems and is based on the sampling with a probe signal sent on the same channel of the qubits. By setting a threshold in the intensity of the probe at the receiver, it is possible to select in real time the lags of high channel transmissivity which correspond to acceptable QBER values.
Indeed, we proved that with the ARTS method we were able to decrease the measured QBER and to extract secret keys in adverse conditions, when the initial average QBER is above the security threshold of 11\%.
         
%
%

\begin{acknowledgements} 
The authors wish to warmly thank for the help provided by Z. Sodnik of the European Space Agency and by S. Ortolani of University of Padova as well as by the Instituto de Astrofisica de Canarias (IAC), and in particular F. Sanchez-Martinez, A. Alonso, C. Warden and J.-C. Perez Arencibia, and by the Isaac Newton Group of Telescopes (ING), and in particular M. Balcells, C. Benn, J. Rey, A. Chopping, and M. Abreu.

This work has been carried out within the Strategic-Research-Project QUINTET of the Department of Information Engineering, University of Padova and the Strategic-Research-Project QuantumFuture (STPD08ZXSJ) of the University of Padova.
\end{acknowledgements}

\end{document}